\documentclass[draftcls,onecolumn]{IEEEtran}
\usepackage{amsfonts}
\usepackage{amssymb}
\usepackage{graphicx}

\begin{document}

\title{A Three-dimensional simulation study of the performance of 
Carbon Nanotube Field Effect Transistors with doped reservoirs and realistic geometry}

%\author{Gianluca Fiori and Giuseppe Iannaccone, MEMBER}
\author{\underline{Gianluca Fiori}$^{1}$, Giuseppe Iannaccone$^{1}$ and Gerhard Klimeck$^2$\\
$^1$ Dipartimento di Ingegneria dell'Informazione, Universit\`a di Pisa, 56100 Pisa, Italy\\
$^2$ School of Electrical Engineering, Network for Computational Nanotechnology,\\
Purdue Universtity, West Lafayette, IN USA.\\
email: {\tt \underline{g.fiori@iet.unipi.it}, Tel. +39 050 2217639}}
\maketitle
\begin{abstract}

In this work, we simulate the expected device performance 
and the scaling perspectives of Carbon nanotube Field Effect Transistors (CNT-FETs),
with doped source and drain extensions. The simulations are
based on the self-consistent solution of the 3D Poisson-Schr\"odinger 
equation with open boundary conditions, within the Non-Equilibrium Green's
Function formalism, where arbitrary gate geometry and device architecture can 
be considered.
The investigation of short channel effects for different
gate configurations and geometry parameters shows that double gate devices offer quasi
ideal subthreshold slope and DIBL without extremely thin gate dielectrics.
Exploration of devices with parallel CNTs show that 
On currents per unit width can be significantly larger than the silicon counterpart,
while high-frequency performance is very promising.

\end{abstract}

\begin{keywords}
Carbon nanotubes, Ballistic transport, 
NEGF, Technology CAD.
\end{keywords}
\IEEEpeerreviewmaketitle
\IEEEoverridecommandlockouts

\section{Introduction}

Carbon nanotubes (CNTs) represent a promising alternative to conventional
silicon technology~\cite{martel} for future nanoelectronics at the end of
the ITRS roadmap~\cite{itrs2003}.
Since the first work on the topic by Iijima~\cite{Ijima}, 
significant improvements  have been achieved, both from the point of view 
of technology and physical modeling.
In particular, Heinze et al.~\cite{Heinze} have demonstrated Schottky Barrier CNT FETs,
where the modulation of the current is mainly determined by the field-induced modulation 
of the nanotube band structure at the CNT ends.  Such a working principle, however, strongly
limits device performance.  The ambipolar behavior, the poor control of the channel,
and the possible degradation of electrical properties like $I_{{\rm on}}/I_{{\rm off}}$ ratio, 
deeply affects such kind of devices, especially nanotubes with large diameters, as shown in~\cite{Appenzeller}. 
To obtain acceptable Drain Induced Barrier Lowering (DIBL), it has been demonstrated 
that downscaling of device dimensions has to follow particular rules, like maintaining the
ratio between the channel length and the oxide thickness larger than 18~\cite{Appenzeller}.

To alleviate these problems different solutions have been proposed
 to achieve channel modulation of the barrier.
Javey et al.~\cite{Javey} and Nosho et al.~\cite{Nosho}
have shown for example that ohmic contacts can be obtained 
by choosing Pd or Ca as metal for the contacts, respectively.
 Inducing charge in the source and drain regions is another possible solution, reported
for example in~\cite{Chen} and in~\cite{Bockrath}.

Channel modulation has also been obtained \cite{Wind,Ming} by the definition
of multiple gates able to fix independent potentials both in the reservoirs and in the channel as well as
transparent Schottky barrier at the contacts.

In a scenario where many geometries are feasible, adequate physical models 
and simulation tools are necessary not only to 
provide explanations to experimental results, but also to define device guidelines 
for the fabrication of CNT FETs, with performance benchmarked against their mainstream silicon
counterpart.
%~\cite{Chow}.

%A simplified approach is that in~\cite{Pourfath}, where 
%the potential profile of a three-dimensional structure is obtained by means of a two-dimensional
% simulator, and transport is computed through the Landauer formalism.

Gate-all-around CNT FETs have been studied~\cite{Clifford1,Clifford2}
where the three-dimensional Poisson equation has been reduced to two dimensions because of the 
cylindrical symmetry of the electrostatic potential, and transport has been computed through the Landauer formalism.
Such coaxial geometry has also been adopted in~\cite{Jing}, where the Poisson equation has been 
coupled with the Non-Equilibrium Green's Function (NEGF)
formalism, using a mode space approach, which enables the computation of transport
 to a small number of electron subbands with a small computational cost.

However, planar gate structures are more attractive because of their simpler fabrication technology,
and all the experimental data discussed above are on planar geometries.

A full three-dimensional approach has been followed in~\cite{Jing2}, 
where the Poisson equation
has been solved using the method of moments. Such a method provides the advantage of
 requiring the computation
of the Poisson equation only in regions where charge is not zero, with the drawback 
 that it is practically impossible to treat more than two different
dielectric constants.

In this work, we focus on realistic and experimentally relevant 
CNT-FETs with doped source and drain extensions and evaluate
their performance against the requirements at the end of the ITRS. 
To this purpose, we have developed a code able to solve the 
 full band Schr\"odinger equations with open boundary conditions in the NEGF framework.
 Such a module has been included in our three-dimensional
Poisson solver NANOTCAD {\sl ViDES}~\cite{ViDES}, which can deal with very
general structures, since it does not take advantage of 
particular symmetries, and can consider structures in which both CNTs and crystalline semiconductors
are simultaneously present. 

Our realistic simulations show that CNT-FETs are very attractive for 
{\it i)} their capabilities of suppressing short channel effects, 
{\it ii)} driving high On currents per unit length 
{\it iii)} their potential for THz applications, providing $I_{on}/I_{off}$ ratio required by the ITRS for the
15~nm gate length, and {\it iv} their potential for THz applications. On the other hand, we shall show that the 
subthreshold slope deteriorates for a gate voltage 
close to zero, due to filling of hole states in the channel.

\section{Simulation Approach}

The potential profile in the three-dimensional simulation domain 
obeys the Poisson equation
\begin{eqnarray}
\nabla \left[\epsilon (\vec{r})\nabla \phi(\vec{r}) \right] & = & -q\left[ p(\vec{r})
-n(\vec{r})+N_D^+(\vec{r}) \right. \nonumber \\
& & \left. -N_A^-(\vec{r}) +\rho_{fix}\right]
\label{Poisson}
,\end{eqnarray}
where $\phi(\vec{r})$ is the electrostatic potential, $\epsilon(\vec{r})$ is the
dielectric constant, $N_D^+$ and $N_A^-$ are the concentration of ionized donors and acceptors, 
respectively and $\rho_{fix}$ is the fixed charge.
The electron and hole concentration ($n$ and $p$, respectively) 
are computed by solving the Schr\"odinger equation with open boundary conditions,
by means of the NEGF formalism~\cite{Dattasuperlattice}. A tight-binding Hamiltonian
with an atomistic (p$_z$ orbitals) real space basis~\cite{Jing3} has been used with 
a hopping parameter $t$=2.7~eV.

The Green's function can then be expressed as
\begin{equation}
G(E)=\left[EI-H-\Sigma_S-\Sigma_D \right] ^ {-1},
\label{green}
\end{equation}
where $E$ is the energy, $I$ the identity matrix, $H$ the Hamiltonian of the 
CNT, and $\Sigma_S$ and $\Sigma_D$ are the self-energies of the source and drain, respectively.
As can be seen, tranport is here assumed to be completely ballistic.

The considered CNTs are all zig-zag nanotubes, 
but the proposed approach can be easily generalized to 
nanotubes with a generic chirality, since the required changes involve only the 
Hamiltonian matrix. Once the length and the chirality of the nanotube are defined, the coordinates in the 
three-dimensional domain of each carbon atom are computed~\cite{Dresselhaus}.
After that, the three-dimensional domain is discretized so that a grid point is defined
in correspondence of each atom, while a user specified grid is defined in regions
not including the CNT. 

A point charge approximation is assumed, i.e. all the free charge around each
carbon atoms is spread with a uniform concentration in the elementary cell including the atom.
In particular, the electron and hole densities are computed from the Density
of States (DOS), derived by the NEGF formalism. Assuming that the chemical potential 
of the reservoirs are aligned at the equilibrium with the flat Fermi level of the CNT, 
and given that are no fully confined states, the electron concentration is
\begin{eqnarray}
n(\vec{r})&=&2\int_{E_i}^{+\infty} dE \left[ {\rm DOS_S(E,\vec{r})} f(E-E_{F_S}) \right. \nonumber \\
&& \left.+ {\rm DOS_D(E,\vec{r})} f(E-E_{F_D}) \right],
\label{equazione}
\end{eqnarray}
while the hole concentration is 
\begin{eqnarray}
p(\vec{r})&=&2\int^{E_i}_{-\infty} dE \left\{ {\rm DOS_S(E,\vec{r})}\left[1- f(E-E_{F_S})\right] \right. \nonumber \\
&& \left.+ {\rm DOS_D(E,\vec{r})} \left[1- f(E-E_{F_D})\right] \right\},
\end{eqnarray}
where $f$ is the Fermi-Dirac occupation factor, and
DOS$_S$ (DOS$_D$) is the density of states injected by the 
source (drain), and E$_{F_S}$ (E$_{F_D}$) is the Fermi level
of the source (drain).

The current has been computed by means of the Landauer formula
\begin{equation}
I=\frac{2q}{h}\int_{-\infty}^{+\infty} dE \mathcal{T}(E) \left[ f(E-E_{F_S}) -f(E-E_{F_D}) \right],
\end{equation} 
where $q$ is the electron charge, $h$ is Planck's constant and $\mathcal{T}(E)$ is the
transmission coefficient computed as
\begin{equation}
\mathcal{T}=-Tr\left[\left(\Sigma_S-\Sigma^\dag_S\right)G\left(\Sigma_D-\Sigma^\dag_D\right)G^\dag\right],
\end{equation}
where $G$ is the Green's function, while $\Sigma_S$ and $\Sigma_D$ are the source and drain 
self-energy matrices~\cite{Dattasuperlattice}.
%\begin{equation}
%\Gamma_{S,D}=i\left[\Sigma_{S,D}-\Sigma^\dag_{S,D}\right].
%\end{equation}
The Green's function is computed  by means of the Recursive Green's Function (RGF)
technique~\cite{Lake97,Anantram}.
A particular attention must be put in the definition of each self-energy matrix, which can be
interpreted as a boundary condition of the Schr\"odinger equation.
In particular, in our simulation we have considered a self-energy for
semi-infinite leads boundary conditions, which enables to consider the CNT 
as connected to infinitely long CNTs at its ends. 

We have to point out that the computation of the self-energy is quite demanding.
In order to achieve faster results, we have followed the approach proposed in~\cite{Rubio}, which
provides results four times faster as compared to a simple under-relaxation method.

We have to point out that, using a real space basis, the computed current
takes into account intra- and inter-band tunneling, since, as compared to the mode space
approach, all the bands of the nanotube are considered simultaneously.

From a numerical point of view, the non-linear system has been solved with the Newton/Raphson (NR)
method with the Gummel iterative scheme.
In particular, the Schr\"odinger equation is solved at the
beginning of each NR cycle of the Poisson equation, 
and the charge density in the CNT is kept constant until
the NR cycle converges (i.e. the correction on the potential is
smaller than a predetermined value).
The algorithm is then repeated cyclically until the norm of 
the difference between the potential computed at the end of two subsequent 
 NR cycles is smaller than a predetermined value.

Some convergence problems however may be encountered using this iterative scheme.
Indeed, since the electron density is independent of the potential 
 within a NR cycle, the Jacobian is null for points of the domain including carbon atoms, loosing
control over the potential correction.
We have then tried , according with the predictor/corrector scheme proposed by Trellakis et al.~\cite{trella}
to find a suitable expression for the charge predictor, in order to give an approximate expression
for the Jacobian at each step of the NR cycle. To this purpose,
we have used an exponential function for the predictor, also found independently by Polizzi et al.~\cite{polizzi}.
In particular, if $n$ is the electron density as in~(\ref{equazione}), the electron density $n_i$ at the $i$-th step
of the NR cycle can be expressed as 
\begin{equation}
n_i=n\exp\left(\frac{\phi_i-\tilde{\phi}}{K_BT} \right),
\end{equation}
where $\tilde{\phi}$ and $\phi_i$ and are the electrostatic potential computed at the first and $i$-th step of 
the NR cycle, respectively, while $K_B$ and $T$ are the Boltzmann constant and the temperature.
Same considerations follow for the hole concentration.
Since the electron density $n$ is extremely sensitive to small changes of the electrostatic potential between
two NR cycles, the exponential function acts in the overall procedure as a dumping factor for charge 
variations.
 In this way, convergence has been improved especially
in the subthreshold regime as well as in the strong inversion regime.
Problems however are still present for example in regions of the device where the charge in the 
nanotube is not compensated by fixed charge, like in the case of bound states in the channel, 
where the right-hand term of the Poisson equation is considerably large.

\section{Results and Discussions}

First, we consider a (11,0) CNT embedded in SiO$_2$, with a diameter $d$ of 0.9~nm, an undoped 
channel of varying length $L$ and n-doped CNT extensions of 
10~nm at the source and drain ends (Fig.~\ref{strutturacnt}). 
The CNT extensions have a stoichiometric ratio $f$ of 
fully ionized donors. 

As a first attempt to study CNT-FET performance, we have considered
the impact of the molar fraction $f$ on the current in the off-state regime.
Fig.~\ref{Irho} shows the current 
for $V_{DS}=0.5~V$ and $V_{GS}=0~V$ as a function of $f$, for two double gate (DG) CNT-FETs
with $t_{ox}$=2 nm, and $L$=7~nm and 15~nm. As can be seen, for the $L$=7~nm device the 
current is extremely sensitive to $f$, and a small variation in the number of 
ionized atoms in the source and drain extensions can determine variations of the
current of almost two orders of magnitude. However, as the channel length is increased, 
such effect is weakened, since the field generated by uncompensated donors in the reservoirs
is less effective in lowering the channel barrier as the channel length is increased.

Since the number of donors is of the order of tens, current dispersion due to random dopant 
fluctuations can be problematic as shown in Fig.~\ref{Irho}, where 
the current for a $L$=7~nm (7,0) CNT is shown.
 In this case, the number of atoms in the reservoirs has been decreased 
by a factor of $\frac{11}{7}$ and the current in the off states varies in this case by almost three orders of magnitude for the same range of doping factor.

For DC properties, we evaluate the devices in terms
of short channel effects, $I_{on}$ and $I_{off}$ currents. 
We first consider short channel effects for different gate layouts 
(single, double and triple gate) for the same channel length 
$L=15$~nm (Fig.~\ref{strutture}).

The subthreshold swing $S$ and DIBL
as a function of the oxide thickness are plotted in Figs.~\ref{SDIBL}a and
 \ref{SDIBL}b. Null Neumann boundary conditions are 
imposed on the lateral faces of the transversal cross sections, in order
to consider an array of CNTs. As expected, the more gates surround the channel, 
the better is channel control. Triple gate devices show an ideal 
behavior even for the thickest oxide we
have considered (5~nm), while 
 quasi-ideal $S$ is obtained for the double gate structure in the whole considered
range of SiO$_2$ thickness. A single gate 
provides acceptable $S$ and DIBL for 2~nm oxides.
Fig.~\ref{SDIBLtox2} shows the $S$ and DIBL
as a function of the channel length for a DG device with oxide thickness $t_{ox}$ of
1 and 2~nm. DG devices show both very good $S$ and DIBL down to 10~nm, and still acceptable 
values at 7~nm, while smaller channel lengths suffer from excessive degradation
of the gate voltage control over the channel.

From here on, we will focus on a double gate structure, 
with $t_{ox}$=2~nm and the cross section shown in Fig~\ref{strutture}b.
Fig.~\ref{Ion}a shows the on-current $I_{on}$ per unit width defined as the
current obtained for $V_{GS}=V_{DS}=0.8~V$, as a function of the channel length,
assuming a lateral dielectric separating adjacent nanotubes of 2~nm.
In Fig.~\ref{Ion}b, $I_{on}$ is plotted as a function
of the nanotube diameter, for a device with channel length equal to 7~nm.

Short channel effects become more 
important, as the channel length is decreased,  and at the same biasing conditions shorter devices 
show larger $I_{on}$ currents, since lowering of the channel barrier 
occurs. Moreover, as far as the CNT diameter is increased,
quantized states along each atom ring are closer in energy so that more subbands
participate to electron transport, increasing channel conductance.

In Fig.~\ref{Ions}a we show the On current as a function of the normalized
tube density per unit length $\rho=d/T$, where $T$ is the distance between the
center of two nanotubes, as illustrated in the inset of Fig.~\ref{Ions}a.
All the results show that CNT-FETs can drive significant currents. As compared
to the ITRS requirements, in the case of the most densely packed 
array, the current per unit length is almost 7 times larger than that
expected for high performance devices at the 32~nm technology node 
(hp32~: effective gate length equal to 13~nm), 
and 6 times for the 22~nm technology node (hp22 : effective gate length equal
to 9~nm). 

The off-current ($I_{off}$), defined as the current obtained for 
$V_{GS}=0~V$ and $V_{DS}=0.8~V$, is 15 times larger than that required 
both for the hp32 and hp22 nodes (Fig.~\ref{Ions}b). 
However, as shown in Fig.~\ref{Ioffspacingnew}a, where the Off-current as
a function of the normalized tube density is depicted, as the density decreases also the 
current in the off-state decreases, so that for tube density smaller than $8\times10^{-2}$
we obtain a $I_{on}/I_{off}$ ratio larger than that required by the ITRS
for the hp32 (Fig.~\ref{Ioffspacingnew}b).

As observed also in~\cite{knoch}, the degradation of the off-current
 is due to the presence of bound states 
in the valence band, which, for high doping and for large 
drain-to-source voltages are occupied
by holes tunneling from the drain reservoir (Fig.~\ref{DOS}a). For smaller
$V_{DS}$, bound states are quite far from the drain Fermi level (Fig.~\ref{DOS}b),
so the linear behavior in the semilog plot of the transfer characteristics in the subthreshold regime
is almost recovered, as shown in Fig.~\ref{idvgs7}.

Another interesting effect due to the bound states, is that in the negative gate voltage regime,
when band-to-band tunneling occurs at both source and drain ends,
 the current increases for negative gate voltage (Fig.~\ref{current}).
For larger $V_{DS}$, this effect requires higher $V_{GS}$ in order to be observed, since
the larger the drain-to-source voltage, the stronger the influence of the bound states in the
valence band, which act against the activation of band-to-band tunneling process.
In addition, as can be seen in the case of $V_{DS}$=0.1~V, for negative gate voltages, resonant
states appear. 
%
%To demonstrate that indeed the observed current oscillations is due to resonant tunneling,
%we plot in Fig.~\ref{resonant}a and Fig.~\ref{resonant}b the density of states in correspondence
%of the second peak, computed for gate voltages
%equal to $V_{GS}$=-2.0~V and $V_{GS}$=-2.1~V respectively, as also hi-lighted in Fig.~\ref{current}.
%Phonon enhanced tunneling will smear out the sharp edges in the I-V~\cite{koswatta}.
%  
%For a gate voltage equal to $V_{GS}$=-2.0~V, the Fermi level of the source is between the two localized
%states, so in the transfer characteristic we observe a decrease of the current, while for 
% $V_{GS}$=-2.1~V the Fermi level of the source is aligned with the second bound state, and electrons
%can tunnel to the drain through the channel. The interesting aspect to be underlined, is that 
%such an effect appear at room temperature, since the confinement in the valence band is so 
%strong that the distance between two bound states is of the order of 150~meV.
%
%
Since in the considered CNTs the band gap is close to 1~eV, the above considerations
suggest that such effect could also limit the performance of silicon devices in the decananometer regime.

Fig.~\ref{gm} shows the transconductance $g_m$ as a function 
of the gate voltage for devices with channel length of 5~nm, 7~nm and 10~nm 
computed for $V_{DS}=0.8$~V.
The transconductance peaks are in correspondence of the gate voltage at which the first one-dimensional 
subband crosses the source Fermi level.
As can be seen, good values of the transconductance are obtained for the all three considered devices. 

We now focus our attention on switching and high-frequency performance of CNTs.
The typical figure of merit for digital applications is 
the intrinsic device speed, defined as 
$\tau = C_GV_{DD}/I_{on}$, where $V_{DD}$ is the 
supplied voltage and $C_G$ is the differential gate capacitance
for $V_{GS}=0.8$~V (Fig.~\ref{tauft}a).
This quantity is typically used to estimate the time it takes an inverter to  
switch, when its output drives another inverter,
represented as a load capacitance $C_G$, as shown in the inset of Fig. \ref{tauft}a.
Compared to the ITRS requirements for the hp22 technology node, the obtained $\tau$  
are  at least 12 times faster.

CNT-FETs also show good potential for THz  applications~\cite{Burke}.
In Fig.~\ref{tauft}b, the cut-off frequency defined as $f_T=\frac{g_m}{2\pi C_G}$
is shown as a function of the channel length~: $f_T$ is 
of the order of tens of THz, and the values obtained by simulations are larger
than those found in~\cite{Burke}, where the gate capacitance is
overestimated. 
As a word of caution, we must consider that additional stray capacitances
could reduce the estimated $f_T$ and $\tau$.

\section*{Conclusion}

We have developed a novel 3D NEGF-based simulation tool
for arbitrary CNT-FET architectures, which 
has enabled us to investigate the performance perspectives of 
CNT-FETs from an engineering point of view.

We have in principle demonstrated that random distribution of dopants in the reservoirs
can significantly affect device performance, and degrade current in the off-state
by several orders of magnitude. We have shown that double-gate structures 
exhibit very small short channel effects even
with rather thick silicon oxide gate dielectric 
(5 nm), and still acceptable subthreshold swing and Drain Induced
Barrier Lowering for devices with channel length down to 7~nm. 

The driving currents and the transconductance
are very good as compared to the ITRS requirements, while
the $I_{off}$ may pose some problems due to the presence 
of localized hole states in the channel. However, good $I_{on}/I_{off}$
can be achieved reducing the tube density in the CNT-FET array,
still satisfying ITRS requirements. 

We have also shown the double gate CNT-FETs are very promising for 
THz applications if stray capacitances can be maintained
under control.\\

\section*{Acknowledgment}

Support from the EU SINANO NoE (contract no. 506844), from
the MIUR-PRIN ``Architectures and models for nanoMOSFETs''
is gratefully acknowledged, and NSF grant \# EEC-0228390.

\newpage

\begin{figure}
\vspace{3cm}
\caption{Three-dimensional structure of the simulated CNT-FETs. }
\label{strutturacnt}

\caption{\label{Irho} Current density for $V_{DS}=0.5~V$ and $V_{GS}=0~V$ 
as a function of the molar
fraction of doping atoms $f$ for Double Gate (11,0) CNT-FETs with $L$=7~nm and $L$=15~nm and
(7,0) $L$=7~nm DG CNT-FET.}

%\caption{\label{Irho7}Current for $V_{DS}=0.5~V$ and $V_{GS}=0~V$ 
%as a function of the molar
%fraction of doping atoms $f$ for Double Gate CNT-FETs with $L$=7~nm and $L$=15~nm. The CNT
%is a (7,0) zig-zag nanotube. }

\caption{\label{strutture} Transversal cross section of the 
CNT-FETs with different gate 
structures : a) single gate; b) double gate; c) triple gate. Null Neumann boundary
condition are imposed on lateral ungated surfaces. }
%\end{figure}

%\begin{figure} 
\caption{\label{SDIBL} a) Subthreshold slope and b) Drain Induced Barrier Lowering 
as a function of the oxide thickness, for $L$=15~nm and for different gate layouts. $f=10^{-3}$. }
%\end{figure}

%\begin{figure} 
\caption{\label{SDIBLtox2} a) Subthreshold slope and b) Drain Induced Barrier Lowering 
as a function of the channel length, for the DG CNT-FETs with $t_{ox}$=2~nm and $t_{ox}$=1~nm .}
%\end{figure}

%\begin{figure} 
\caption{\label{Ion} a) On Current per unit width as a function of the channel length for a double
 gate CNT-FET (2~nm lateral dielectric between adjacent nanotubes); 
b) $I_{on}$ current per nanotube as a function of the nanotube 
diameter, for a $L$=7~nm double gate CNT-FET. $t_{ox}=2$~nm, $f=5\times10^{-3}$. }
%\end{figure}

%\begin{figure} 
\caption{\label{Ions} a) On current as a function of
the nanotube normalized density per unit length $\rho=d/T$ for a double gate CNT-FET with $L$=15~nm;
b) Off-current 
as a function of the channel length for a double gate CNT-FET.}
%\end{figure}

%\caption{\label{Ioff} Off-current 
%as a function of the channel length for a double gate CNT-FET.}

\caption{\label{Ioffspacingnew} a) Off-current as a function of
the nanotube normalized density per unit length $\rho=d/T$ for a double gate CNT-FET with $L$=15~nm;
b) $I_{on}/I_{off}$ ratio as a function of
the nanotube normalized density per unit length $\rho=d/T$ for a double gate CNT-FET with $L$=15~nm.}

%\caption{\label{Ionoffratio} $I_{on}/I_{off}$ ratio as a function of
%the nanotube normalized density per unit length $\rho=\frac{d}{T}$ for a $L$=15~nm double gate CNT-FET.}

\caption{\label{DOS}Density of states computed for the device with $L$=7~nm, for $V_{GS}=0~V$ 
and a) $V_{DS}=0.8~V$, b) $V_{DS}=0.5~V$ as a function of the energy and
the coordinate along the nanotube axis. Dashed lines are in correspondence
of the source and drain Fermi level.}

\caption{\label{idvgs7} Transfer characteristics for the double gate 
CNT-FET  with $L$=7~nm, for 
$V_{DS}=0.5~V$ and $V_{DS}=0.8~V$; $t_{ox}=2$~nm, $f=5\times10^{-3}$.}

\caption{\label{current} Transfer characteristics for the double gate 
CNT-FET  with $L$=7~nm, for 
$V_{DS}=0.5~V$ and $V_{DS}=0.1~V$; $t_{ox}=1$~nm, $f=5\times10^{-3}$.  }

%\caption{\label{resonant} Density of states for the double gate CNT-FET  with $L$=7~nm, for 
%$V_{DS}=0.1~V$, a) $V_{GS}=-2.0~V$ and b) $V_{GS}=-2.1~V$. $t_{ox}=1$~nm, and $f=5\times10^{-3}$.  }

\caption{\label{gm} Transconductance as a function of the gate voltage 
for double gate CNT-FETs with different channel lengths : $L$=5~nm, 7~nm and 10~nm;
  $t_{ox}=2$~nm, $V_{DS}=0.8$~V, and $f=5\times10^{-3}$.}

\caption{\label{tauft} a) Inverse of the intrinsic device speed, defined as 
$\tau=C_GV_{DD}/I_{on}$ as a function of the 
channel length for double gate CNT-FET; $V_{DD}=0.8V$,
$C_G$ is the gate capacitance. b) Cut-off frequency as a function of the gate length, for the 
double gate CNT-FET.  $t_{ox}=2$~nm,$f=5\times10^{-3}$.}

\end{figure}

\newpage

\begin{figure}
\vspace{3cm}
\begin{center}
\includegraphics[width=9.5cm]{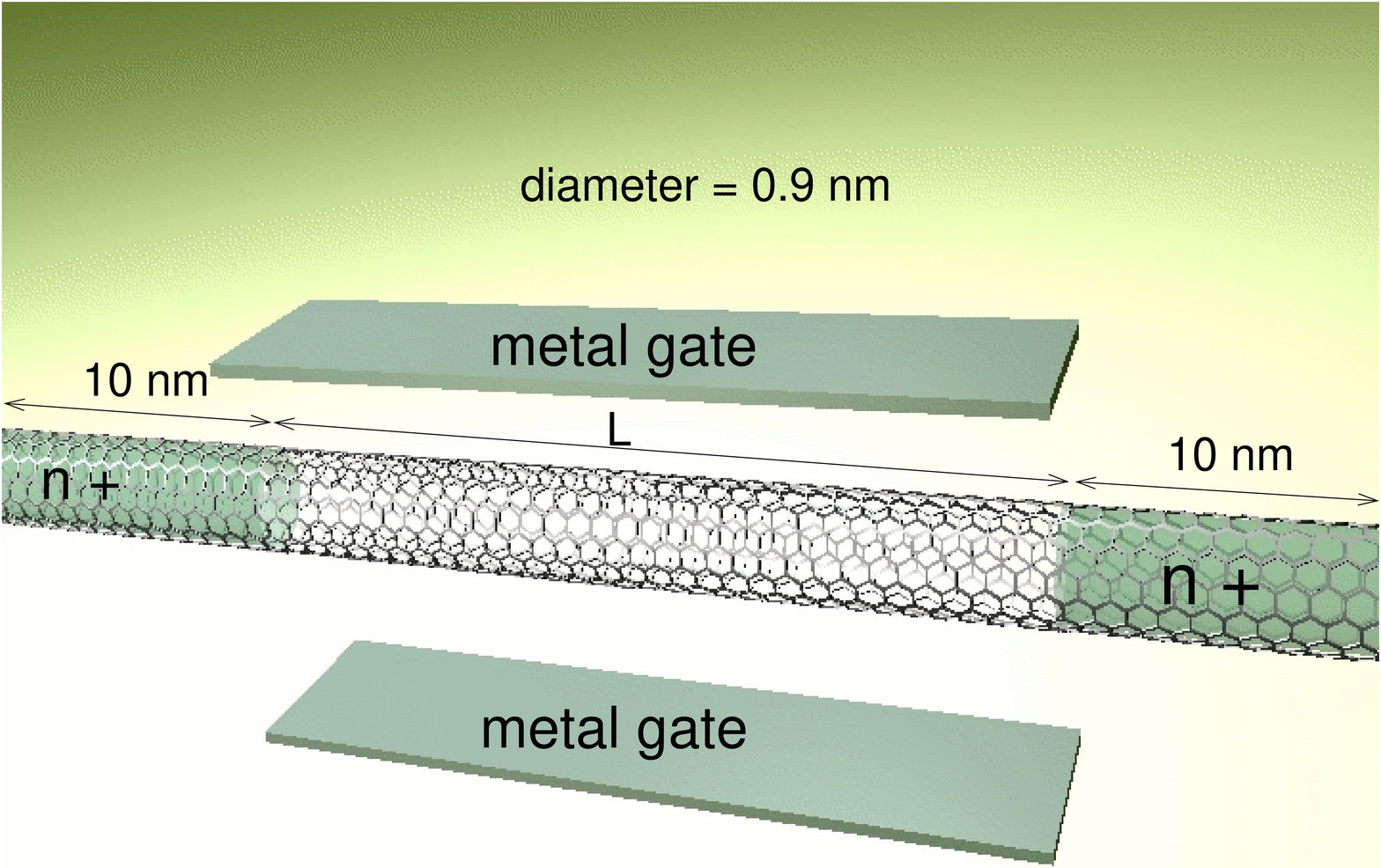}
\huge
\\
FIG. 1 \\
Gianluca Fiori et al.\\
\normalsize
\end{center}
\end{figure}

\newpage

\begin{figure}
\vspace{3cm}
\begin{center}
\includegraphics[width=11.5cm]{./Irhofuno.eps}% Here is how to import EPS art
\huge
\\
FIG. 2 \\
Gianluca Fiori et al.\\
\normalsize
\end{center}
\end{figure}

%\newpage
%
%\begin{figure}
%\vspace{3cm}
%\begin{center}
%\includegraphics[width=9.5cm]{./Irhonew7,0.eps}% Here is how to import EPS art
%\huge
%\\
%FIG. 3 \\
%Gianluca Fiori, Giuseppe Iannaccone and Gerhard Klimeck\\
%IEEE Trans. on  Electron Devices\\
%\normalsize
%\end{center}
%\end{figure}

\newpage

\begin{figure}
\vspace{3cm}
\begin{center}
\includegraphics[width=9.5cm]{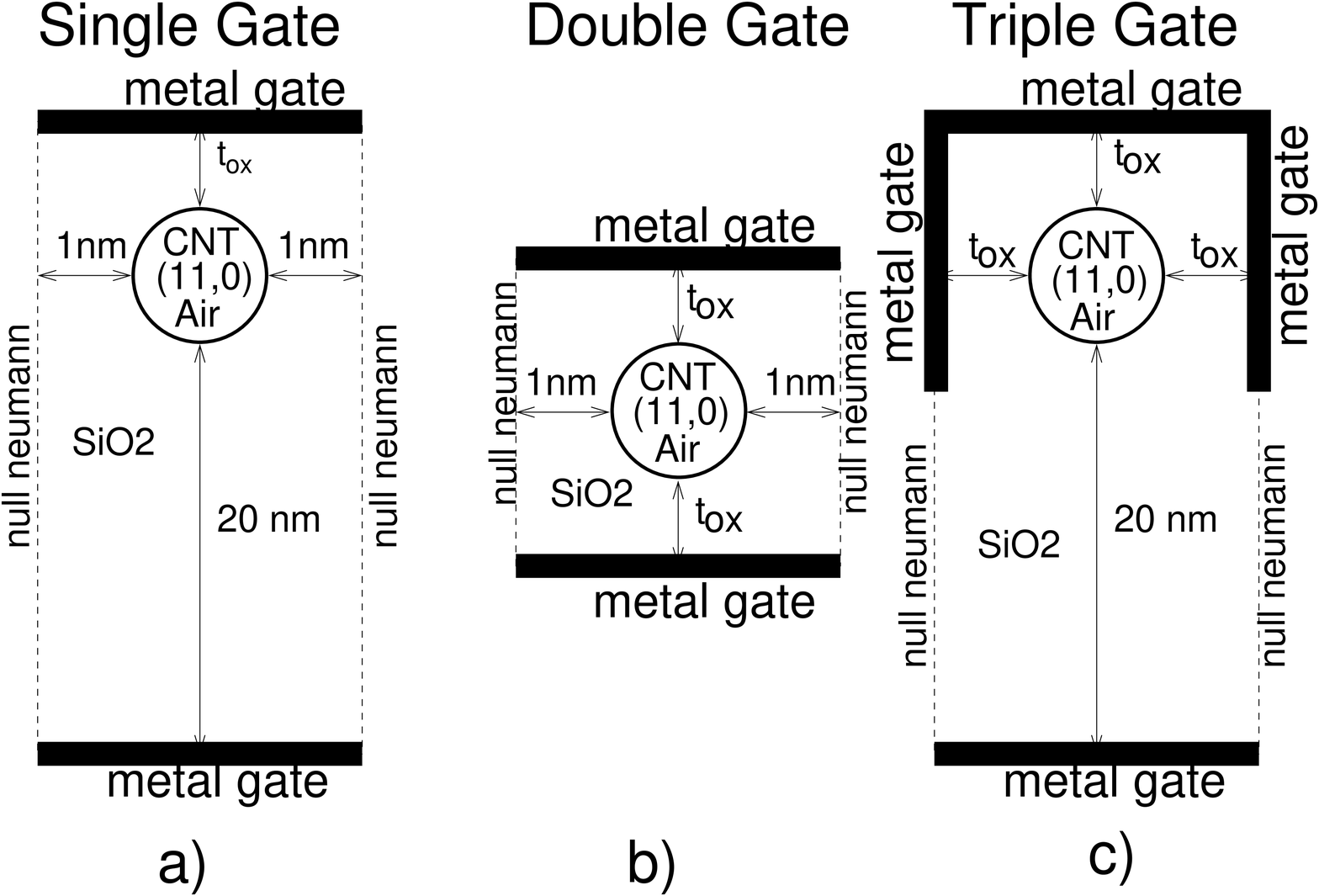}
\huge
\\
FIG. 3 \\
Gianluca Fiori et al.\\
\normalsize
\end{center}
\end{figure}

\newpage

\begin{figure}
\vspace{3cm}
\begin{center}
\includegraphics[width=11.5cm]{./SDIBL.eps}
\huge
\\
FIG. 4 \\
Gianluca Fiori et al.\\
\normalsize
\end{center}
\end{figure}

\newpage

\begin{figure}
\vspace{3cm}
\begin{center}
\includegraphics[width=11.5cm]{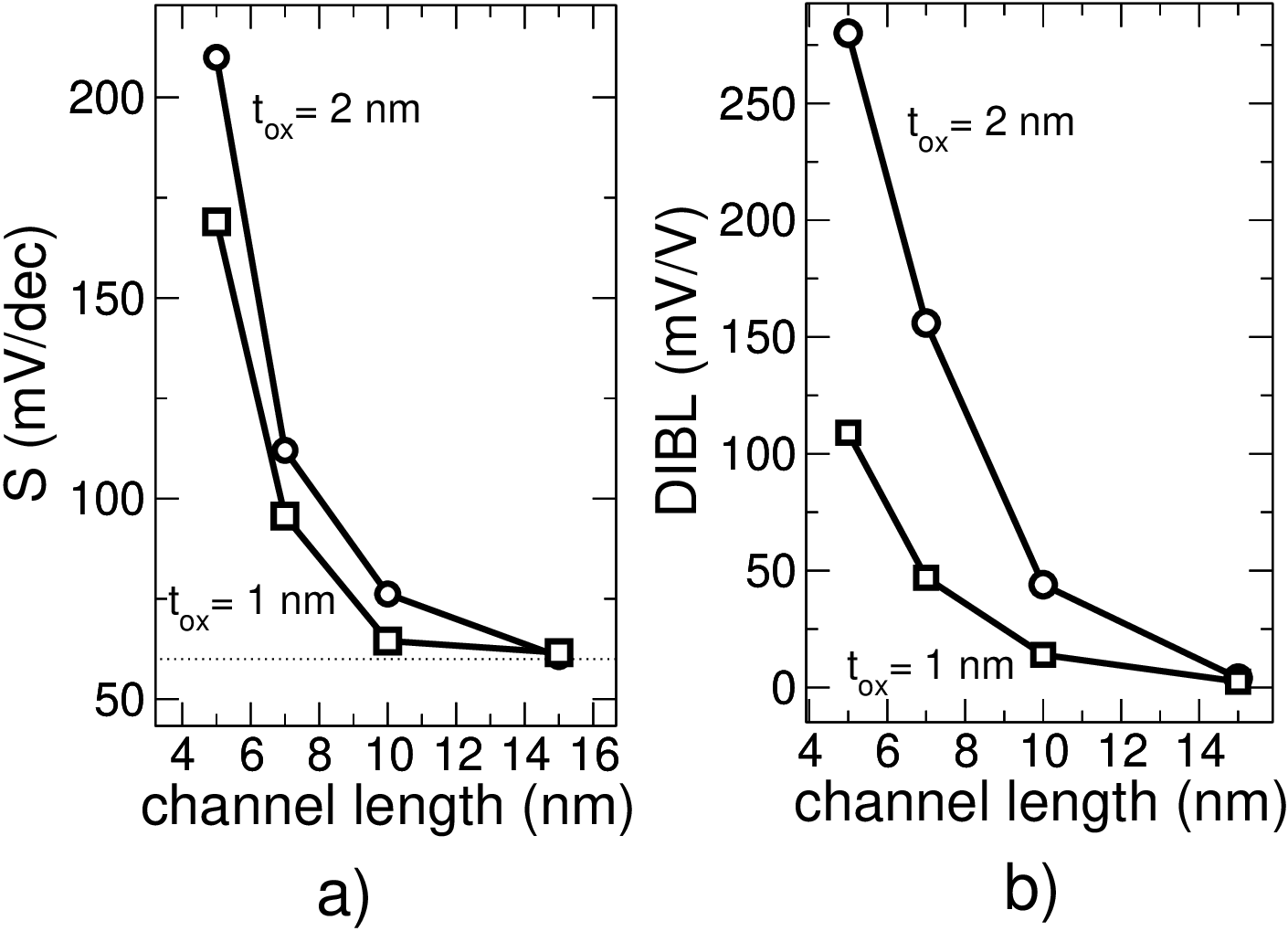}
\huge
\\
FIG. 5 \\
Gianluca Fiori et al.\\
\normalsize
\end{center}
\end{figure}

\newpage

\begin{figure}
\vspace{3cm}
\begin{center}
\includegraphics[width=11.5cm]{./Ionnewplusdiam2.eps}
\huge
\\
FIG. 6 \\
Gianluca Fiori et al.\\
\normalsize
\end{center}
\end{figure}

\newpage

\begin{figure}
\vspace{3cm}
\begin{center}
\includegraphics[width=11.5cm]{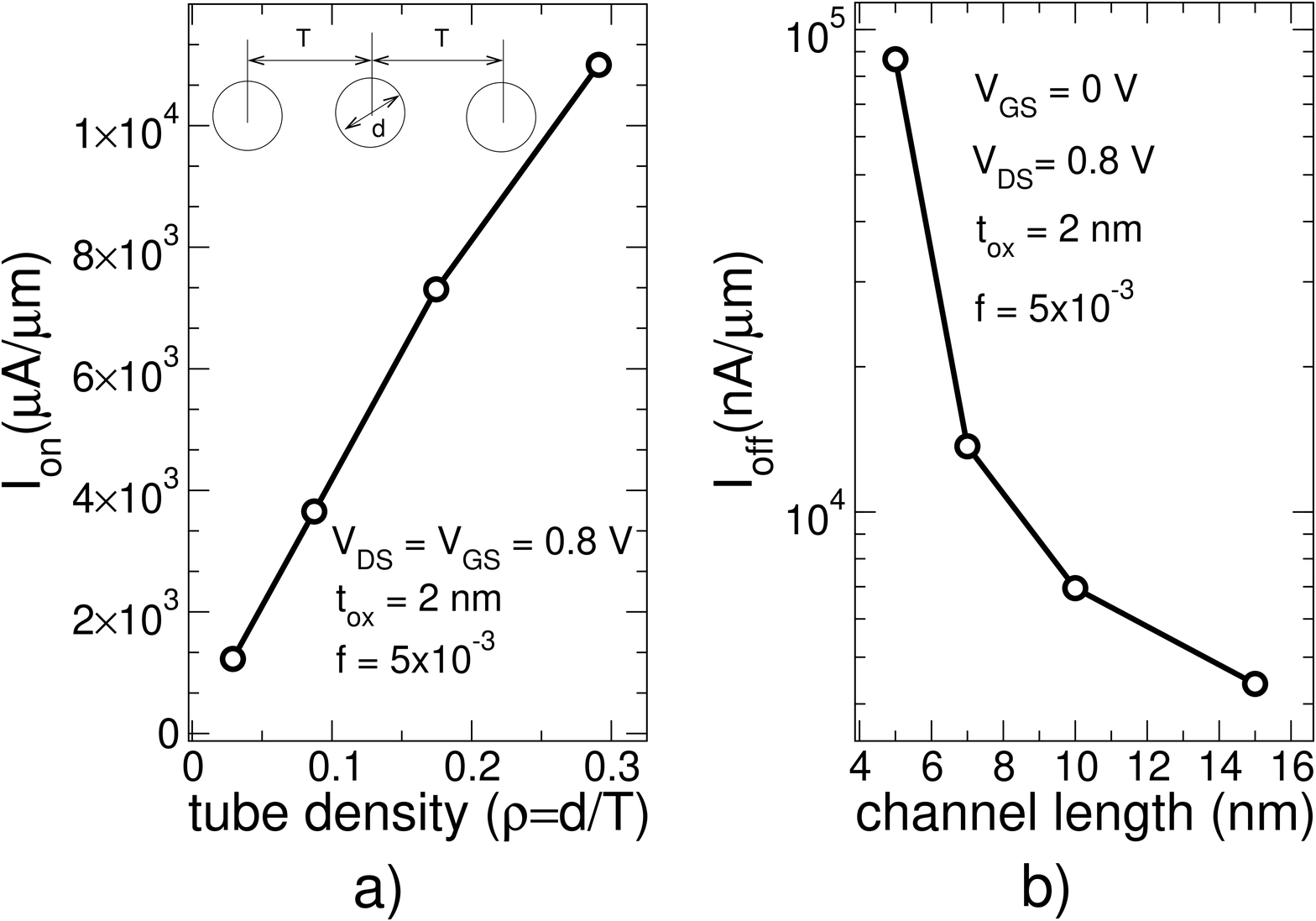}
\huge
\\
FIG. 7 \\
Gianluca Fiori et al.\\
\normalsize
\end{center}
\end{figure}

%\newpage
%
%\begin{figure}
%\vspace{3cm}
%\begin{center}
%\includegraphics[width=9.5cm]{./Ioffnew.eps}
%\huge
%\\
%FIG. 8 \\
%Gianluca Fiori, Giuseppe Iannaccone and Gerhard Klimeck\\
%IEEE Trans. on  Electron Devices\\
%\normalsize
%\end{center}
%\end{figure}

\newpage

\begin{figure}
\vspace{3cm}
\begin{center}
\includegraphics[width=11.5cm]{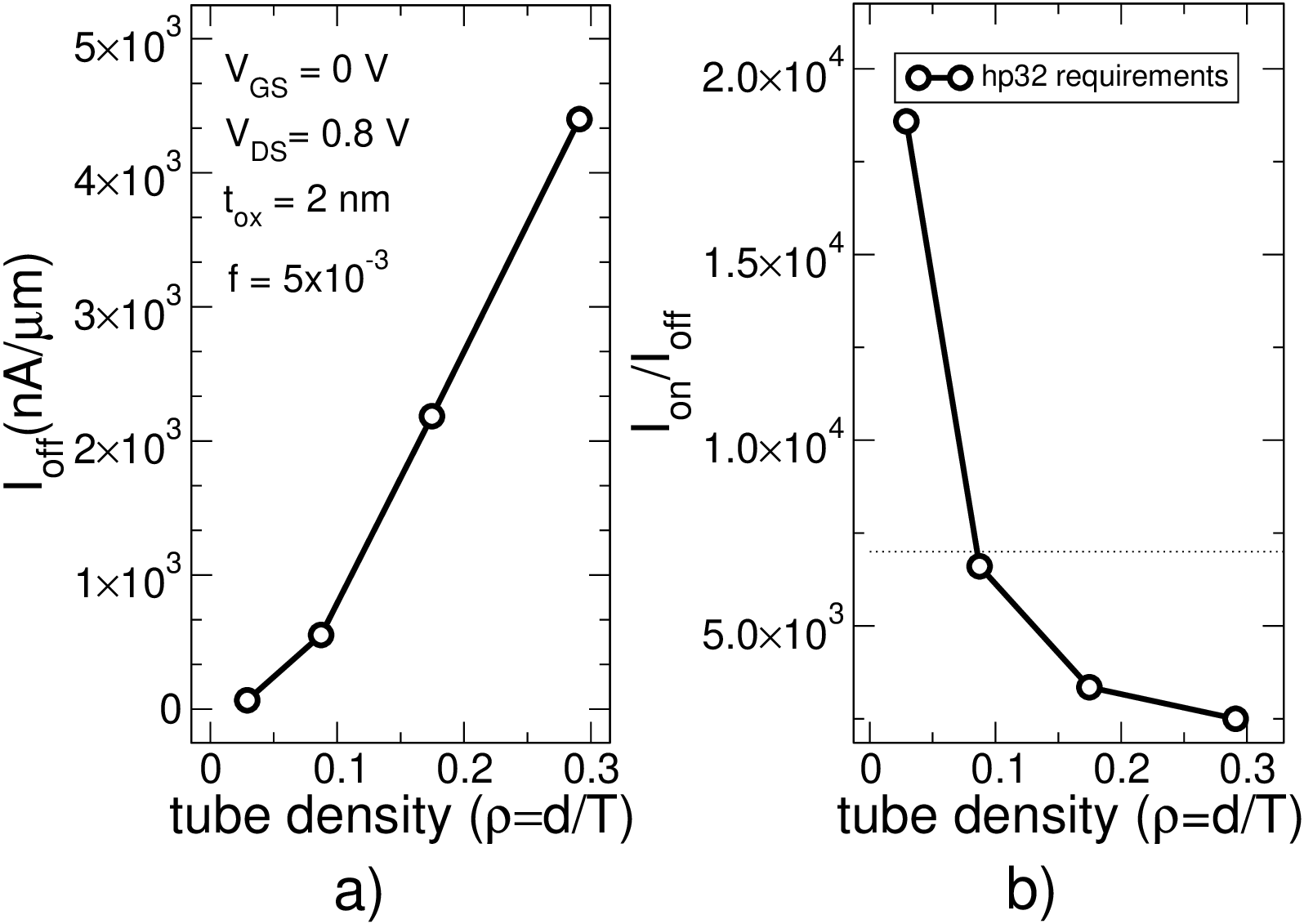}
\huge
\\
FIG. 8 \\
Gianluca Fiori et al.\\
\normalsize
\end{center}
\end{figure}

%\newpage
%
%\begin{figure}
%\vspace{3cm}
%\begin{center}
%\includegraphics[width=9.5cm]{./Ionoffratio.eps}
%\huge
%\\
%FIG. 9 \\
%Gianluca Fiori, Giuseppe Iannaccone and Gerhard Klimeck\\
%IEEE Trans. on  Electron Devices\\
%\normalsize
%\end{center}
%\end{figure}

\newpage

\begin{figure}
\vspace{3cm}
\begin{center}
\includegraphics[width=9.5cm]{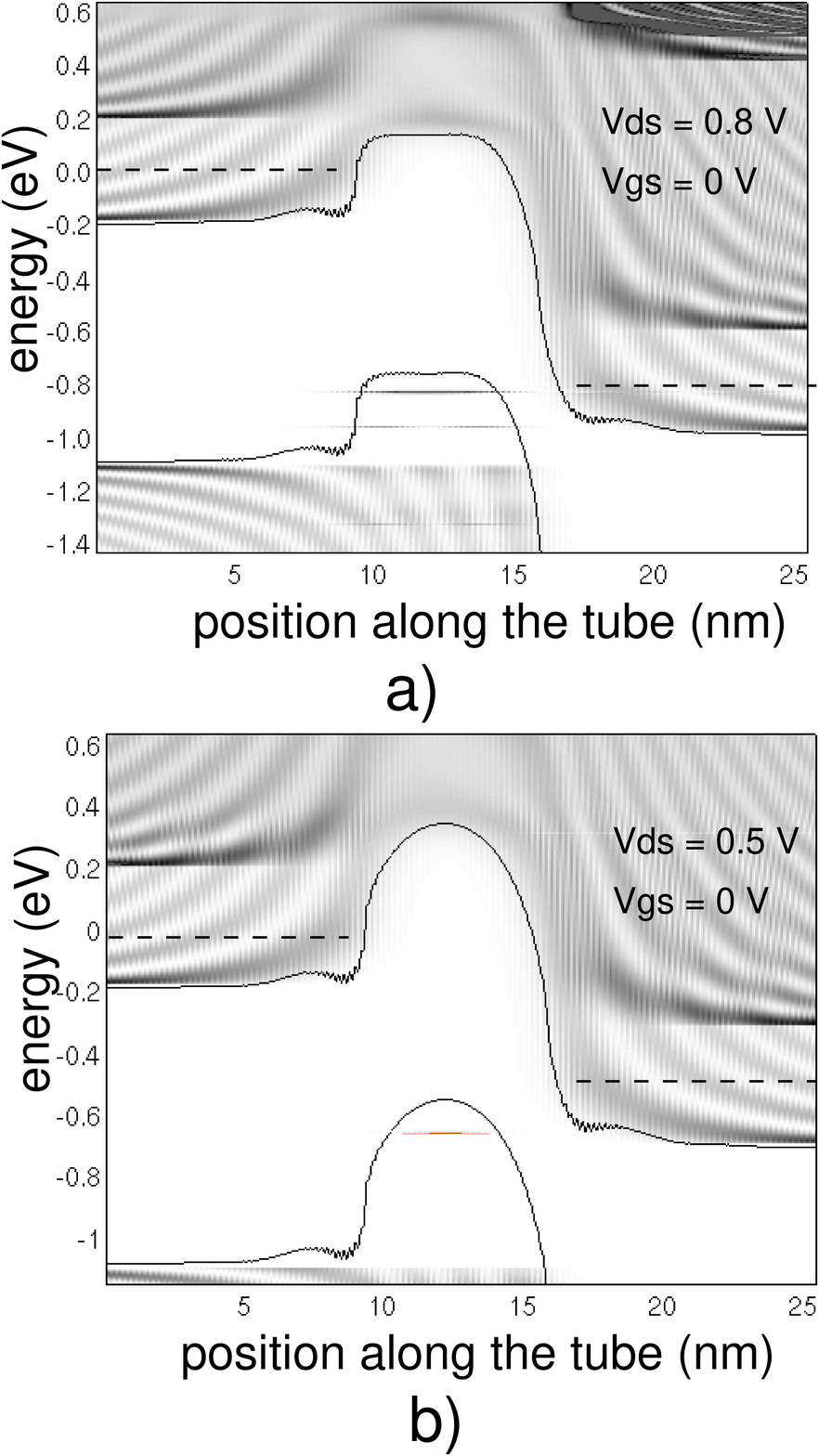}
\huge
\\
FIG. 9 \\
Gianluca Fiori et al.\\
\normalsize
\end{center}
\end{figure}

\newpage

\begin{figure}
\vspace{3cm}
\begin{center}
\includegraphics[width=9.5cm]{./idvgs7new.eps}
\huge
\\
FIG. 10 \\
Gianluca Fiori et al.\\
\normalsize
\end{center}
\end{figure}

\newpage

\begin{figure}
\vspace{3cm}
\begin{center}
\includegraphics[width=9.5cm]{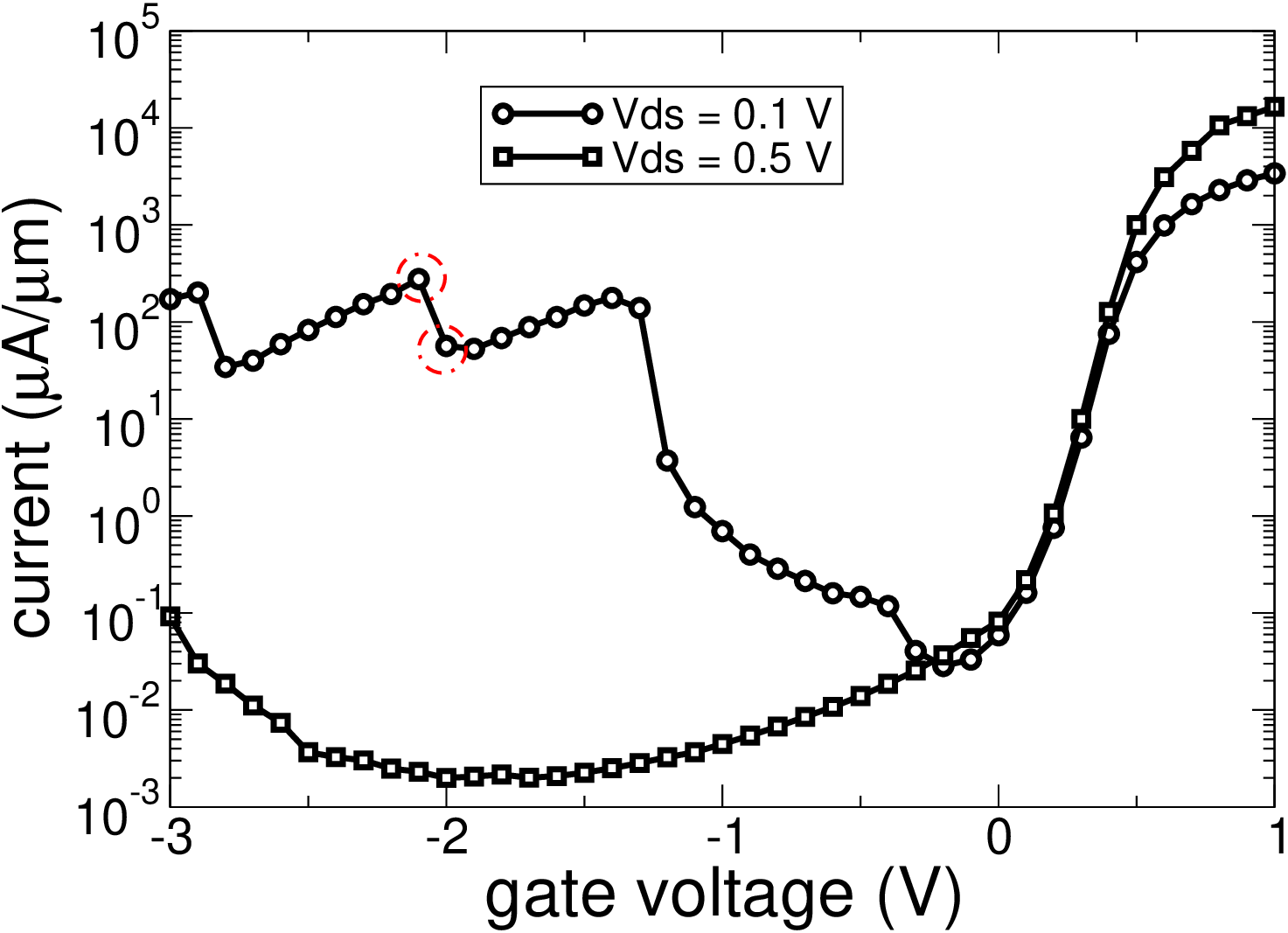}
\huge
\\
FIG. 11 \\
Gianluca Fiori et al.\\
\normalsize
\end{center}
\end{figure}

%\newpage
%
%\begin{figure}
%\vspace{3cm}
%\begin{center}
%\includegraphics[width=9.5cm]{./DOSresonant.eps}
%\huge
%\\
%FIG. 12 \\
%Gianluca Fiori et al.\\
%IEEE Trans. on  Electron Devices\\
%\normalsize
%\end{center}
%\end{figure}
%
\newpage

\begin{figure}
\vspace{3cm}
\begin{center}
\includegraphics[width=9.5cm]{./gmnew2.eps}
\huge
\\
FIG. 12 \\
Gianluca Fiori et al.\\
\normalsize
\end{center}
\end{figure}

\newpage

\begin{figure}
\vspace{3cm}
\begin{center}
\includegraphics[width=9.5cm]{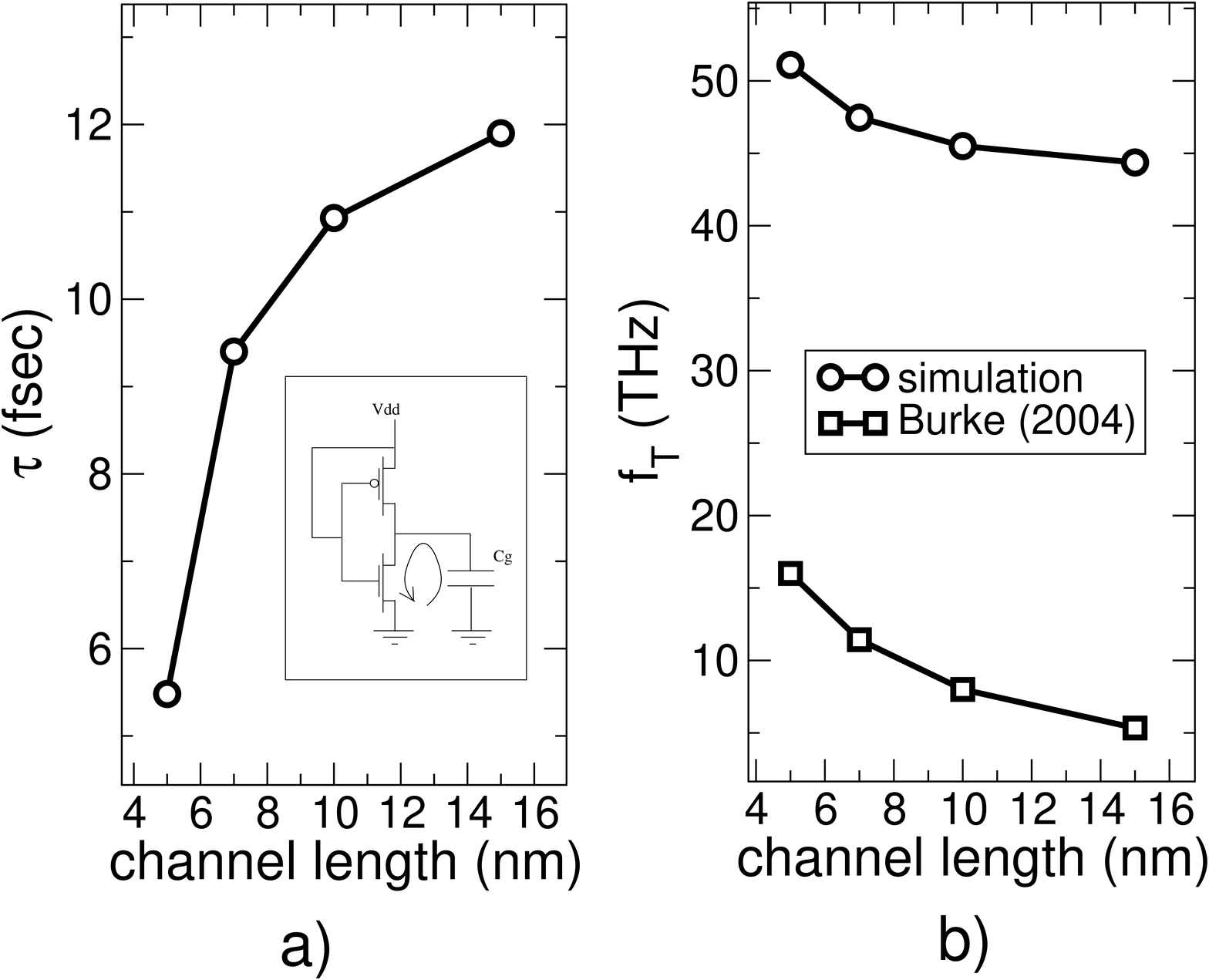}
\huge
\\
FIG. 13 \\
Gianluca Fiori et al.\\
\normalsize
\end{center}
\end{figure}

\end{document}